\documentclass[12pt,a4paper]{article}

\usepackage{amsmath,amsthm,amssymb,amsfonts}
\usepackage[a4paper,text={450pt,650pt},centering]{geometry}

\makeatletter
\let\@keywords\@empty
\let\@subject\@empty
\providecommand{\keywords}[1]{\gdef\@keywords{#1}}
\providecommand{\subject}[1]{\gdef\@subject{#1}}
\providecommand{\address}[1]{\gdef\@address{#1}}
\providecommand{\email}[1]{\gdef\@email{#1}}
\def\thetitle{\@title}
\def\theauthor{\@author}
\def\thesubject{\@subject}
\def\theaddress{\@address}
\def\theemail{\@email}
\def\thedate{\@date}
\def\thekeywords{\@keywords}
\makeatother

\def\nnn{\nonumber \\}
\def\eref{\eqref}

\def\be{\begin{equation}}
\def\ee{\end{equation}}

\def\lb{\bigl[}
\def\rb{\bigr]}

\def\lp{\bigl(}
\def\rp{\bigr)}
\def\lcurl{\bigl\{}
\def\rcurl{\bigr\}}

\def\a{\alpha}
\def\b{\beta}
\def\cm{{\cal M}}
\def\ck{{\cal K}}
\def\d{\delta}
\def\D{\Delta}
\def\hc{\hat{c}}
\def\hd{\hat{d}}
\def\k{\kappa}
\def\l{\lambda}
\def\m{\mu}
\def\n{\nu}

\def\O{\Omega}
\def\p{\phi}
\def\vp{\varphi}

\def\t{\tau}
\def\pa{\partial}

\def\Thp{\Theta_{+}}
\def\Thm{\Theta_{-}}
\def\Thpm{\Theta_{\pm}}

\def\lie{{\cal G}}
\def\alie{{\widehat \lie}}

\def\Complex{\mathbb{C}}
\def\Integer{\mathbb{Z}}

\def\H#1{H^{#1}}
\def\Ea#1{E_{\alpha}^{#1}}
\def\Eb#1{E_{-\alpha}^{#1}}

\def\D#1{D^{(#1)}}
\def\Thp{\Theta_{+}}
\def\Thm{\Theta_{-}}

\def\tpsi{\tilde{\psi}}
\def\tchi{\tilde{\chi}}
\def\bra#1{\langle #1|}
\def\ket#1{|#1\rangle}

\def\inner#1#2{\langle #1 | #2 \rangle}

\def\NPB#1#2#3{{\sl Nucl. Phys.} {\bf B#1} (#2) #3}

\def\CMP#1#2#3{{\sl Commun. Math. Phys.} {\bf #1} (#2) #3}

\def\JMP#1#2#3{{\sl J. Math. Phys.} {\bf #1} (#2) #3}

\def\IJMPA#1#2#3{{\sl Int. J. Mod. Phys.} {\bf A#1} (#2) #3}

\def\JPA#1#2#3{{\sl J. Physics} {\bf A#1} (#2) #3}

\begin{document}

\linespread{1.2}

\title{Dressing approach to the nonvanishing boundary 
value problem for the AKNS hierarchy}

\author{J. F. Gomes, Guilherme S. Fran\c ca and A. H. Zimerman}

\address{
Instituto de F\'\i sica Te\' orica - IFT/UNESP\\
Rua Dr. Bento Teobaldo Ferraz, 271, Bloco II\\
01140-070, S\~ ao Paulo - SP, Brazil}

\email{[jfg,guisf,zimerman]@ift.unesp.br}

\keywords{dressing, vertex operator, boundary condition, integrable models}
\subject{MSC (2010): 17B67, 81U15}


\begin{titlepage}
\begin{center}
{\Large\bfseries\thetitle\par}\vspace{1cm}
\textsc{\theauthor}\par\vspace{5mm}
\textit{\theaddress}\par\vspace{3mm}
\texttt{\theemail}\par\vspace{1cm}
November 29, 2011\par\vspace{1cm}
\textbf{Abstract}\par\vspace{7mm}
\begin{minipage}{\textwidth}
We propose an approach to the nonvanishing boundary value
problem for integrable hierarchies based on the dressing method. Then
we apply the method to the AKNS hierarchy. The solutions are
found by introducing appropriate vertex operators that takes into
account the boundary conditions.
\end{minipage}
\end{center}
\end{titlepage}

\section{Introduction}
\label{sec:intro}

The structure of affine Kac-Moody algebras has provided considerable 
insight underlining the construction of integrable hierarchies in 1+1 
spacetime dimensions. An integrable hierarchy consists of a series of 
equations of motion with a common algebraic structure.
They are constructed out of a zero curvature  condition which involves a 
pair of gauge potentials, lying  within the affine Lie algebra. A crucial 
ingredient is the decomposition of the affine Lie algebra $\alie$ into 
graded subspaces, determined by a grading operator $Q$. The integrable 
hierarchy is further specified by a constant, grade one semi-simple 
element $E^{(1)}$. The choice of $\alie$, $Q$ and $E^{(1)}$ are the basic 
ingredients that fully determine and classify the entire hierarchy  
(see for instance ref. \cite{nissimov} for a review).

Representation theory of Kac-Moody algebras play also an important role 
in constructing systematically soliton solutions of integrable hierarchies.  
The main idea of the dressing method consists in, by gauge transformation, 
to map a simple vacuum configuration into a non trivial one- or multi-soliton 
solution. The  pure gauge solution of the zero curvature representation 
leads to explicit spacetime dependence for the vacuum configuration which 
in turn, generates by gauge transformation, the non-trivial soliton solutions 
of the hierarchy  \cite{babelon, mira}.  It becomes  clear that the
solutions are classified into  conjugacy classes according to the  choice of 
vacuum configuration.  In general, the vacuum is taken as the  zero field 
configuration and the soliton solutions are constructed and classified in 
terms of vertex operators \cite{olive}.  

A dressing method approach
to nontrivial constant vacuum configuration was proposed in 
\cite{mkdv_even} when considering the negative even flows of the 
modified Korteweg-de Vries (mKdV) hierarchy, which do not admit trivial 
vacuum configuration, i.e. 
vanishing boundary condition. 
A deformed vertex operator was introduced and the method was further 
applied to the whole mKdV hierarchy with nonvanishing boundary condition 
and to a hierarchy containing 
the Gardner equation \cite{gardner}. Here in 
these notes, we shall provide further clarifications  
discussing nonvanishing boundary conditions for integrable hierarchies in 
general, constructed with the structure proposed in \cite{jfg} 
(and references therein). We point out
that not all the models within the hierarchy admit solutions with nonvanishing
(constant) boundary condition. We then consider as an example the 
Ablowitz-Kaup-Newell-Segur (AKNS) hierarchy with nontrivial constant vacuum 
configuration. 
Results for the focusing nonlinear Schroedinger (NLS) equation were 
recently obtained in ref. \cite{zakharov}. 
A hybrid of the dressing and
Hirota method for a multi-component generalization of the AKNS hierarchy 
was also considered recently in \cite{blas}. 

We first  describe the general aspects of the dressing method in 
section~\ref{sec:dressing}, together with a discussion on the possible 
different boundary conditions. We point out the requirements for the individual
models of the hierarchy to admit a nonvanishing constant boundary 
value solutions.
Later in section~\ref{sec:akns}, we discuss the construction of the AKNS 
hierarchy, followed by the construction of its solutions, 
section~\ref{sec:solutions}.
In section~\ref{sec:vertex} we present the deformed vertex operators used to
construct solitonic solutions with nonvanishing boundary condition and
illustrate with explicit examples.

\section{Algebraic concepts and integrable hierarchies}
\label{sec:algebra}

In this section we introduce the
main algebraic concepts used for the construction of integrable models and
to obtain its solutions. A more detailed exposition can be found 
in \cite{leznov, goodard}.

Let $\lie$ be a semi-simple finite-dimensional Lie algebra. 
The infinite-dimensional loop algebra $L(\lie)$ is defined as the tensor 
product of $\lie$ with integer powers of the so called 
complex spectral parameter $\l$, 
$L(\lie)\equiv\lie\otimes\Complex[\l,\l^{-1}]$.
If $T_a\in\lie$ then $T_a^{n}\equiv T_a\otimes\l^n \in L(\lie)$ for 
$n\in\Integer$. The commutator of the loop algebra is given by
\be
\lb T_a^{n},T_b^{m} \rb \equiv \lb T_a,T_b \rb\otimes\l^{n+m},
\ee
where $\lb T_a,T_b\rb$ is the $\lie$ commutator.

The central extension is performed by the introduction of the operator 
$\hc$, which commutes with all the others $\lb \hc, T_a^{n}\rb=0$. 
Furthermore, consider the spectral derivative operator 
$\hd\equiv\l\tfrac{d}{d\l}$ such
that $\lb \hd, T_a^{n}\rb=nT_a^{n}$, so it measures the \emph{power} of the 
spectral parameter. The affine Kac-Moody Lie algebra
is defined by $\alie\equiv L(\lie)\oplus\Complex\hc\oplus\Complex\hd$ and
for $T_a^{n},T_b^{m}\in \alie$ the commutator is provided by
\be
\lb T_a^{n}, T_b^{m} \rb \equiv \lb T_a, T_b\rb\otimes\l^{n+m}+
n\delta_{n+m,0}\inner{T_a}{T_b}\hc,
\ee
where $\inner{T_a}{T_b}$ is the Killing form on $\lie$.

The algebra $\alie$ can be decomposed into $\Integer$-graded subspaces by
the introduction of a grading operator $Q$ such that
\be\label{grading}
\alie = \bigoplus_{j\in\Integer}\alie^{(j)},
\ee
where 
$\alie^{(j)}\equiv\lcurl T^{n}_a \in \alie \ \vert\ 
\lb Q,T_a^{n}\rb=j T_a^{n}\rcurl$. The integer $j$ is the \emph{grade} 
of the operators defined with respect to $Q$. It follows from the Jacobi 
identity
that if $T_a^{n}\in\alie^{(i)}$ and $T_b^{m}\in\alie^{(j)}$ then
$\lb T_a^{n}, T_b^{m}\rb \in \alie^{(i+j)}$.

Let $E$ be a semi-simple element of $\alie$. Its kernel  is 
defined by 
$\ck\equiv\lcurl T_a^{n} \in \alie \ \vert \ \lb E, T_a^{n}\rb=0 \rcurl$.
Its complement is called the image subspace $\cm$ and 
$\alie=\ck\oplus\cm$. It can be readily verified from this definition and
the Jacobi identity that $\lb \ck, \ck \rb \subset \ck$ and 
$\lb\ck, \cm\rb\subset \cm$. We also assume the
symmetric space structure $\lb \cm, \cm \rb\subset \ck$.

Consider the linear system in 1+1 spacetime dimension
\be
\lp \pa_x + U \rp\Psi=0, \qquad \lp \pa_t + V \rp\Psi=0, 
\ee
where $U, V \in \alie$ and $\Psi$ is an element of the Lie 
group of $\alie$. The compatibility of this
system ensures integrability, yielding the zero curvature equation
\be\label{zerocurv}
\lb \pa_x + U, \pa_t + V \rb = 0.
\ee
The general structure to construct nontrivial integrable models through an 
algebraic
approach was discussed in \cite{jfg, leznov} (see also  references therein).
Let $Q$ be a grading operator
and choose a constant, independent of $x$ and $t$, grade one semi-simple 
element $E=E^{(1)}$. Let
$A^{(0)}(x,t)$ be a linear combination of zero grade operators lying entirely
in $\cm$. The coefficients of this linear combination
are spacetime functions, i.e. the fields of the model. Therefore, if
$U$ and $V$ have the following algebraic structure \cite{jfg}
\begin{subequations}\label{construction}
\begin{align}
U &\equiv E^{(1)}+A^{(0)}(x,t), \label{construction_u} \\
V &\equiv \sum_{i=-p}^{q}D^{(i)}(x,t) = D^{(q)}+D^{(q-1)}+\dotsb+D^{(-p)},
\label{construction_v}
\end{align}
\end{subequations}
where $p,q \ge 0$ are nonnegative integers, the zero curvature 
equation \eref{zerocurv} can be solved grade by grade, determining
each $D^{(i)}$ in terms of the fields in $A^{(0)}(x,t)$; the zero
grade projection will provide the nonlinear integrable
field equations. The explicit examples
considered in the following sections will illustrate this assertion.
The particular case where $V=D^{(-1)}$ is 
equivalent to a relativistic model obtained from a Hamiltonian reduction of the
Wess-Zumino-Novikov-Witten model. The infinite number differential equations
obtained from this algebraic structure, one for each choice of $p$ and $q$,
constitute an integrable hierarchy immediately classified by 
$\alie$, $Q$ and $E^{(1)}$.

Consider a unitary highest weight representation of $\alie$, where the
hermitian relations are $\lp T_a^{n}\rp^{\dagger}=T_a^{-n}$. There exists
a finite set of states $\lcurl \ket{\mu_k}\rcurl$, 
$k=0,1,\ldots,\textnormal{rank}\hat {\lie}$, 
that are annihilated by every operator with a positive power on 
the spectral parameter,
\be\label{states}
T_a^n\ket{\mu_k} = 0, \qquad n > 0.
\ee
The fundamental state $\ket{\mu_0}$ is annihilated by
every operator, including $\lcurl T_a^0 \rcurl$, except the central term.
Under the action of $\hc$ all states obey
\be\label{fundamental}
\hc \ket{\mu_k} = \ket{\mu_k}.
\ee

\section{Dressing method}
\label{sec:dressing}

Let us briefly review the dressing method \cite{babelon, mira}.
Assume a known solution of the zero curvature equation \eref{zerocurv}, 
in pure gauge form, called vacuum
\be\label{vacuum}
U_0 = -\pa_x\Psi_0\Psi_0^{-1}, \qquad
V_0 = -\pa_t\Psi_0\Psi_0^{-1}.
\ee
The dressing method arises as a factorization problem
\be\label{dressing}
\Thm^{-1}\Thp = \Psi_0 g \Psi_0^{-1},
\ee
where $g$ is an arbitrary constant group element, independent of $x$ and $t$, 
and $\Thpm$ are factorized into positive and negative grade operators, 
respectively, according to a grading operator \eref{grading},
\be\label{gauss_grade}
\Thm = \exp\lp m^{(-1)}\rp\exp\lp m^{(-2)}\rp \dotsm, \qquad
\Thp = \exp\lp m^{(0)}\rp\exp\lp m^{(1)}\rp \dotsm.
\ee
where $m^{(j)} \in \alie^{(j)}$.
The factorization problem \eqref{dressing} ensures that, the 
dressing operators $\Thpm$ map a vacuum solution
into a new solution by the gauge transformations
\begin{subequations}\label{gauge}
\begin{align}
U&=\Thpm U_0\Thpm^{-1}-\pa_x\Thpm\Thpm^{-1},\label{gauge1}\\
V&=\Thpm V_0\Thpm^{-1} - \pa_t\Thpm\Thpm^{-1}.\label{gauge2}
\end{align}
\end{subequations}

\subsection{Vanishing boundary condition}
\label{sec:vc}

Taking a zero field configuration $A^{(0)}\to 0$ in \eref{construction}
we have the constant vacuum operators
\begin{subequations}\label{zero_vac}
\begin{align}
U_0 &= E^{(1)} \equiv \O^{(1)}, \label{zero_vac_u}\\
V_0 &= D^{(q)}_0 + D_0^{(-p)} \equiv \O^{(q)} + \O^{(-p)}. \label{zero_vac_v}
\end{align}
\end{subequations}
where $\O^{(j)}\in\ck$. This conclusion comes from the projection of the zero
curvature equation into each graded subspace and the resulting equations for
the highest and lowest grades \cite{jfg}. 
Using \eqref{zero_vac_u} in the gauge transformation
\eqref{gauge1} with $\Thp$, taking into account the central extension%
\footnote{In the construction of integrable models only the loop algebra
plays a significant role, but to employ the highest weight representation 
the central term needs to be carefully considered. The equations of motion
are invariant under $U \to U + f \hc$. We are choosing $f=-\nu_x$ for
a convenient notation.}
of the loop algebra denoted by $-\nu_x\hc$, and projecting the resulting
equation into a zero graded subspace we get
\be\label{gauge_zero}
A^{(0)} - \nu_x\hc = - \pa_x\exp\lp m^{(0)}\rp \exp\lp-m^{(0)}\rp.
\ee
Assume a pure gauge connection, involving zero grade generators only, 
for the field operator
\be\label{ansatz_zero}
A^{(0)}=-B_xB^{-1}.
\ee
Substituting \eqref{ansatz_zero} into \eqref{gauge_zero} gives
\be\label{g0_zero}
\exp\lp m^{(0)}\rp = B \exp\lp\nu\hc\rp.
\ee
Consider the fundamental state $\ket{\mu_0}$ and 
generic states $\ket{\mu_k}$, $\ket{\mu_l}$ that are not annihilated
by some operators ${T_a^{0}}$. In fact, the generic states are chosen
as the ones that survive under the action of the zero grade operators 
associated with the fields in $A^{(0)}$. This will be clear when we consider
a concrete example.
Therefore, projecting \eqref{dressing}
between appropriate highest weight states, we get the implicit solution 
of the model
\be\label{solution_zero}
\bra{\mu_k} B \ket{\mu_l} = \frac{\tau_{kl}}{\tau_{00}},
\ee
where we have defined the tau functions
\be\label{tau_func}
\tau_{00} \equiv \bra{\mu_0}\Psi_0 g \Psi_0^{-1}\ket{\mu_0},\qquad
\tau_{kl} \equiv \bra{\mu_k}\Psi_0 g \Psi_0^{-1}\ket{\mu_l}.
\ee

The vacuum operators \eqref{zero_vac} will commute,
since for zero field configuration they lie in $\ck$,
so integrating \eqref{vacuum} yields
\be\label{psi_zero}
\Psi_0 = \exp\lcurl-\O^{(1)}x-\lp \O^{(q)}+\O^{(-p)}\rp t \rcurl.
\ee
Soliton solutions arise by choosing $g$ in terms of vertex operators 
in the form \cite{olive}
\be\label{g_soliton}
g = \prod_{i=1}^{N}\exp\lp X_i\rp.
\ee
Suppose the vertex is an eigenstate of the vacuum, i.e.
\be\label{eigen}
\lb \O^{(1)}x + \lp \O^{(q)}+\O^{(-p)}\rp t, X_i \rb = -\eta_i(x,t)X_i.
\ee
The function $\eta_i$ contains the dispersion relation in an explicit
spacetime dependence. Then,
the tau functions \eref{tau_func} can be computed in an explicit way,
\be
\Psi_0g\Psi_0^{-1}=\Psi_0\left(\prod_{i=1}^{N} e^{X_i}\right) \Psi_0^{-1}
=\prod_{i=1}^{N}\Psi_0 e^{X_i} \Psi_0^{-1}
=\prod_{i=1}^{N} e^{\Psi_0 X_i \Psi_0^{-1}}.
\ee
Using the expansion $e^{A}Be^{-A}=B+[A,B]+\tfrac{1}{2!}[A,[A,B]]+\dotsb$ with
the eigenvalue equation \eref{eigen} we obtain 
the $N$-soliton solution
\be\label{tau_soliton}
\begin{split}
\t_{kl} &= \bra{\mu_k}\prod_{i=1}^{N}\exp\lp e^{\eta_i} X_i\rp\ket{\mu_l}
   = \bra{\mu_k}\prod_{i=1}^{N}\lp 1+ e^{\eta_i} X_i \rp\ket{\mu_l} \\
   &=\delta_{kl}+a_1 e^{\eta_1}+a_2e^{\eta_2}+a_3e^{\eta_3}\\
&\qquad+a_{12} e^{\eta_1+\eta_2}+ a_{13} e^{\eta_1+\eta_3}+ 
a_{23}e^{\eta_2+\eta_3}+a_{123} e^{\eta_1+\eta_2+\eta_3}+\dotsb,
\end{split}
\ee
where we have assumed the nilpotency property of the vertex,
$\bra{\mu_k}\lp{X_i}\rp^{n}\ket{\mu_l}=0$ for $n\ge2$, and defined the 
following matrix elements
\be
a_i = \bra{\m_k} X_i \ket{\m_l}, \qquad
a_{ij} = \bra{\m_k} X_iX_j \ket{\m_l}, \qquad
a_{ijk} = \bra{\m_k} X_iX_jX_k \ket{\m_l}, \qquad \text{etc.}
\ee

\subsection{Nonvanishing boundary condition}
\label{sec:nvc}

Let us now assume a constant, nonzero, field configuration 
$A^{(0)}\to A^{(0)}_0$.
Then \eqref{construction} yields the constant vacuum operators
\begin{subequations}\label{const_vac}
\begin{align}
U_0 &= E^{(1)}+A^{(0)}_0, \label{const_vac_u} \\
V_0 &= D^{(n)}_0 + D^{(n-1)}_0 + \dotsm. \label{const_vac_v}
\end{align}
\end{subequations}
The analogous of eq. \eqref{gauge_zero} becomes
\be\label{gauge_const}
A^{(0)} - \nu_x\hc = \exp\lp m^{(0)}\rp A^{(0)}_0 \exp\lp-m^{(0)}\rp -
\pa_x\exp\lp m^{(0)}\rp \exp\lp-m^{(0)}\rp.
\ee
Assuming  the form \eqref{g0_zero}, then \eqref{gauge_const} yields a 
relation between the physical fields in $A^{(0)}$, functionals associated
with zero grade generators in $B$ and the nonzero vacuum, i.e.
\be\label{fields_const}
A^{(0)} = B A^{(0)}_0 B^{-1} - B_xB^{-1}.
\ee
When $A^{(0)}_0\to 0$ we recover the relation \eqref{ansatz_zero}.
The general solution of the model is still given in the form
\be\label{solution_const}
\bra{\mu_k} B \ket{\mu_l} = \frac{\tau_{kl}}{\tau_{00}}=
\frac{\bra{\mu_k}\Psi_0 g \Psi_0^{-1}\ket{\mu_l}}
{\bra{\mu_0}\Psi_0 g \Psi_0^{-1}\ket{\mu_0}}
\ee
but now with $\Psi_0$ satisfying the linear system 
\eqref{vacuum} with the more general vacuum 
operators \eref{const_vac} and henceforth, depending 
upon the constant vacuum solution.
We emphasize that the implicit
solution \eqref{solution_const} and \eqref{fields_const} is valid for an
arbitrary vacuum solution, even if it is a general spacetime 
function. For the general case, however, the difficulty arises in integrating 
the linear system \eqref{vacuum}, while for a constant vacuum solution this 
is trivial.

Assuming a constant vacuum, the operators \eqref{const_vac} does 
not lie purely in $\ck$ anymore, but they are supposed to be solution of the
zero curvature equation, i.e. $\lb U_0, V_0 \rb=0$, so integrating
\eqref{vacuum} gives
\be\label{vacuum_sol}
\Psi_0 = \exp\lp-U_0 x - V_0 t\rp.
\ee
Assuming again that $g$ has the form \eqref{g_soliton} and that there exists
vertex operators satisfying eigenvalue equations like \eqref{eigen}, but
now with operators \eqref{const_vac}, it will
be possible to compute the tau functions in \eqref{solution_const}.
Because \eqref{const_vac} depend
on the constant vacuum fields it suggests that for \eqref{eigen}
to be valid, $X_i$ will also depend on these constants fields, so
the dispersion relation and multi-solitonic interaction terms will 
depend on the boundary conditions.

\subsection{Models admitting nonvanishing boundary value solutions}
\label{sec:admitting}

Not all the models in the hierarchy defined by \eqref{construction}
have solutions with nonvanishing boundary 
condition. The vacuum must be a solution of the
zero curvature equation, but for this to be true, in the case
of constant boundary condition, a restriction
on the structure of the Lax pair is required. 

The operator \eqref{construction_v} depends on 
the fields. When taking the fields to vanish, only the 
operator with highest (and/or lowest) grade that is a constant operator 
purely in 
$\ck$ remains, see \eqref{zero_vac_v}. Clearly, in this situation there
is no restriction besides the own algebraic construction 
\eqref{construction}.

For a nonzero constant field configuration, other
operators in \eqref{const_vac_v} that previously vanished now remain, and
the whole combination should commute with \eqref{const_vac_u}.
In \eqref{const_vac_u} we have the structure 
$\O^{(1)}\equiv\ck^{(1)}+\cm^{(0)}$. If \eqref{const_vac_v} keeps the
same algebraic structure, i.e. if it is formed by a linear combination of 
operators of the kind 
\be\label{kind}
\O^{(j)}\equiv\ck^{(j)}+\cm^{(j-1)}=\lambda^{j-1}\O^{(1)}
\ee
where $j\le n$ and $\l$ is the loop algebra spectral parameter, then we will 
have $\lb U_0, V_0 \rb=0$. Note that in \eqref{kind} it is required at 
least two operators differing by one grade. This will not be possible for 
the first model of the negative
hierarchy, $V=\sum_{i=-p}^{-1}D^{(i)}$, that contains only one operator 
$V=D^{(-1)}$. These are
the relativistic models of the hierarchy, like sinh-Gordon and 
Lund-Regge for instance. In sum, at least 
the relativistic models of any hierarchy constructed with the structure 
\eqref{construction} will not have solutions with nonvanishing (constant) 
boundary condition.

\section{The AKNS hierarchy}
\label{sec:akns}

The previously discussed approach for a nontrivial vacuum solution was 
considered
for the mKdV hierarchy in \cite{mkdv_even, gardner}. Now we present another 
example by considering the non-abelian AKNS hierarchy.
Let $\alie=\hat{A}_{1}\sim\hat{s\ell}_2$ with the following commutation
relations
\be
\lb \H{n}, \H{m} \rb = 2n\d_{n+m,0}\hc, \quad 
\lb \H{n}, E_{\pm\a}^{m}\rb = \pm 2 E_{\pm \a}^{n+m}, \quad 
\lb \Ea{n}, \Eb{m} \rb = \H{n+m} + n\d_{n+m, 0}\hc
\ee
Consider the homogeneous grading operator
$Q=\hd$, which decomposes the algebra into the graded subspaces
\be\label{grading_akns}
\alie^{(j)} = \lcurl \H{j},\;\Ea{j},\;\Eb{j} \rcurl.
\ee
Choosing the constant semi-simple element in \eref{construction_u} as
$E^{(1)}=\H{1}$, we have $A^{(0)}=q\Ea{0}+r\Eb{0}$ 
where $q=q(x,t)$ and $r=r(x,t)$ are both fields of the models within the
hierarchy.
Let
\be\label{dj_akns}
\D{j}=a_j \H{j}+b_j \Ea{j}+c_j \Eb{j} \ \in \ \alie^{(j)}
\ee
where $a_j,b_j,c_j$ are functions of $x$ and $t$.
Neglecting for the moment the central terms (i.e.  $\hat c = 0$) and 
considering only positive grade operators in \eqref{construction_v}
we have the zero curvature equation for the positive AKNS
hierarchy
\be\label{akns_hierarchy}
\lb \pa_x+\H{1}+q\Ea{0}+r\Eb{0},
\pa_t+\D{n}+\D{n-1}+\cdots+\D{0} \rb = 0,
\ee
where $n=1,2,3,\ldots$ is an arbitrary positive integer.
Solving \eqref{akns_hierarchy} grade by grade, starting from the 
highest one, we can determine all 
coefficients in \eqref{dj_akns} in terms of the fields $q$ and $r$.
The zero grade projection provides the field equations. 

For the case where $n=2$, fixing 
the two arbitrary constant coefficients as $a_2=1$ and $a_1=0$, and 
ignoring possible integration constants as usual, we get the 
famous AKNS system
\be\label{akns-sist}
q_t = -\tfrac{1}{2}q_{xx}+q^2 r, \qquad
r_t = \tfrac{1}{2}r_{xx}-qr^2.
\ee
Choosing $q=\vp$, $r=\sigma \vp^*$, $t\to-it$ and $x\to ix$ we obtain the well 
known NLS equation
\be
i\vp_t - \tfrac{1}{2}\vp_{xx} - \sigma |\vp|^2\vp = 0.
\ee
The cases $\sigma=\pm 1$ correspond to the \emph{focusing} and 
\emph{defocusing} NLS equations, respectively.
Note that the nonzero constant fields $q\to q_0$, $r\to r_0$ are 
not solution of \eqref{akns-sist}. If we do not ignore an integration
constant arising when solving for the $a_0$ coefficient in \eqref{dj_akns},
but denote it conveniently as $\tfrac{1}{2}q_0r_0$, we get the
system
\be\label{akns-nvc}
q_t =-\tfrac{1}{2}q_{xx}+\lp qr-q_0r_0\rp q, \qquad 
r_t =\tfrac{1}{2}r_{xx}-\lp qr-q_0r_0\rp r,
\ee
whose Lax pair is given by
\begin{subequations}\label{nls_lax}
\begin{align}
U &= \H{1} + q\Ea{0} + r \Eb{0}, \label{u_akns} \\
V &= \H{2} + q\Ea{1} + r\Eb{1}
-\tfrac{1}{2}\lp qr-q_0r_0\rp \H{0}-\tfrac{1}{2}q_x\Ea{0}
+\tfrac{1}{2}r_x\Eb{0}. \label{v_akns2}
\end{align}
\end{subequations}
Note that now $q\to q_0$, $r\to r_0$ is a solution of \eqref{akns-nvc}.
The choice $q=\vp$, $r=\sigma\vp^*$, $t\to-it$ and $x\to ix$ yields the  
\emph{focusing} and \emph{defocusing} NLS equations admitting 
nonvanishing (constant) boundary value solution
\be
i\vp_t - \tfrac{1}{2}\vp_{xx}-\sigma\lp|\vp|^2-A^2\rp = 0,
\ee
where $A^2 \equiv |\vp_0|^2$ is a real constant. 
Solutions of this equation 
with $\sigma=1$ (focusing) was the object under study in the recent paper 
of Zakharov and Gelash \cite{zakharov}.

Solving \eqref{akns_hierarchy} for the $n=3$ case we get
\be\label{akns3}
q_t = \tfrac{1}{4}q_{xxx}-\tfrac{3}{2}qrq_x, \qquad
r_t = \tfrac{1}{4}r_{xxx}-\tfrac{3}{2}qrr_x,
\ee
whose Lax pair given by \eqref{u_akns} and
\begin{align}
V &= \H{3}+q\Ea{2}+r\Eb{2}-\tfrac{1}{2}qr\H{1}-\tfrac{1}{2}q_x\Ea{1}+
\tfrac{1}{2}r_x\Eb{1} \nnn
&\qquad
+\tfrac{1}{4}\lp rq_x-qr_x\rp \H{0}+\tfrac{1}{4}\lp q_{xx}-2q^2r\rp \Ea{0}+
\tfrac{1}{4}\lp r_{xx}-2q^2r\rp \Eb{0}. \label{v_akns3}
\end{align}
The constant fields $q\to q_0$, $r\to r_0$ are solution
of \eqref{akns3}. Note that taking $r=\pm q$ in \eqref{akns3} yields
the mKdV equation, $r=\pm1$ the KdV equation and $r=\a + \b q$ the
Gardner equation with arbitrary coefficients
\be
4 q_t = q_{xxx} -6\a q q_x - 6\b q^2 q_x.
\ee
Solutions with nonvanishing boundary condition for the mKdV 
hierarchy and solutions of a deformed hierarchy containing the Gardner 
equation were considered in \cite{gardner}.

The first negative flow of the AKNS hierarchy is the Lund-Regge model, that
as discussed in section~\ref{sec:admitting}, does not have 
nonvanishing (constant) boundary value solutions.

\section{AKNS solutions}
\label{sec:solutions}

Let 
\be\label{omega_n}
\O^{(n)} \equiv \H{n} + q_0\Ea{n-1} + r_0\Eb{n-1}.
\ee
Taking $q \to q_0$, $r\to r_0$ in \eqref{nls_lax} and
\eqref{v_akns3} yield the respective vacuum Lax operators
\begin{subequations}\label{akns_vacuum}
\begin{align}
U_0 &= \O^{(1)}& &\textnormal{for \eref{u_akns}},\\ 
V_0 &= \O^{(2)}& &\textnormal{for \eref{v_akns2}},\\
V_0 &= \O^{(3)}-\tfrac{1}{2}q_0r_0\O^{(1)}&  
&\textnormal{for \eref{v_akns3}}.
\end{align}
\end{subequations}
Let
\be\label{b_fields}
B = \exp\lp \chi\Eb{0}\rp \exp\lp \phi\H{0}\rp \exp\lp \psi\Ea{0}\rp.
\ee
Solving \eqref{fields_const} will imply in the following relations
\be\label{fields_nvc}
\begin{split}
\phi_x &= r_0\psi\lp 1+\chi\psi e^{2\phi}\rp -q_0\chi e^{2\phi}+
\chi\psi_x e^{2\phi}, \\
q &= e^{2\phi}\lp q_0-r_0\psi^2-\psi_x\rp, \\
r &= r_0 e^{-2\phi}+e^{2\phi}\lp q_0\chi^2-r_0\chi^2\psi^2\rp -\chi_x-
\chi^2\psi_x e^{2\phi}. 
\end{split}
\ee
Introducing the auxiliary functions
\be\label{aux_fields}
\tchi \equiv \chi e^{\phi}, \qquad 
\tpsi \equiv \psi e^{\phi}, \qquad
\Delta \equiv 1+\tchi\tpsi,
\ee
the expressions \eqref{fields_nvc} can be further simplified to
\be\label{qr_aux}
q = \frac{q_0}{\Delta}e^{2\phi}-\frac{\tpsi_x}{\Delta}e^{\phi}, \qquad
r = r_0\Delta e^{-2\phi}-\tchi_x e^{-\phi}.
\ee
Taking into account the central terms (i.e. $\hat c \neq 0 $),  
consider the highest weight states of $\hat{s\ell}_2$; 
$\lcurl \ket{\mu_0}, \, \ket{\mu_1}\rcurl$. These states obey the 
following actions,
\begin{gather}
\H{n}\ket{\m_k}=0, \qquad E_{\pm\a}^{n}\ket{\m_k}=0, \qquad 
\textnormal{(for $n>0$)} \\
E_{\a}^{0}\ket{\m_k}=0, \qquad \H{0}\ket{\m_0}=0, \qquad
\H{0}\ket{\m_1}=\ket{\m_1}, \qquad \hc\ket{\m_k}=\ket{\m_k}.
\end{gather}
The adjoint relations are
\be
\lp \H{n}\rp^{\dagger}=\H{-n}, \qquad 
\lp E_{\pm\a}^{n}\rp^{\dagger}=E_{\mp\a}^{-n}, \qquad 
\hc^{\dagger} = \hc, \qquad \hd^{\dagger}=\hd.
\ee
Besides the highest weight states we use the same notation to denote
the particular state 
\be
\ket{\m_2}=\Eb{0}\ket{\mu_1}\ne0.
\ee
Using these states in \eqref{solution_const} 
with \eqref{b_fields}, we have the following tau functions
\be\label{taufunc}
\begin{split}
\t_{00} &= e^{\n}=\bra{\m_0}\Psi_0g\Psi_0^{-1}\ket{\m_0}, \\
\t_{11} &= e^{\n+\p}=\bra{\m_1}\Psi_0g\Psi_0^{-1}\ket{\m_1}, \\
\t_{12} &= \psi e^{\n+\p}=\bra{\m_1}\Psi_0g\Psi_0^{-1}\ket{\m_2}, \\
\t_{21} &= \chi e^{\n+\p}=\bra{\m_2}\Psi_0g\Psi_0^{-1}\ket{\m_1},
\end{split}
\ee
from which we have
\be\label{fields_tau}
e^{\phi} = \frac{\tau_{11}}{\tau_{00}},\qquad
\chi = \frac{\tau_{21}}{\tau_{11}},\qquad
\psi = \frac{\tau_{12}}{\tau_{11}},\qquad
\tchi = \frac{\tau_{21}}{\tau_{00}},\qquad
\tpsi = \frac{\tau_{12}}{\tau_{00}}.
\ee
Substituting the tau functions in \eqref{qr_aux}
we get the explicit dependence for the nonvanishing boundary
value solutions of the AKNS hierarchy
\begin{subequations}\label{akns_fields_tau}
\begin{align}
q &= q_0\frac{\t_{11}^2}{\t_{00}^2+\t_{12}\t_{21}}+
\frac{\t_{11}}{\t_{00}}\frac{\lp\t_{00}\rp_x\t_{12}-\t_{00}\lp\t_{12}\rp_x}
{\t_{00}^2+\t_{12}\t_{21}},\\
r &= r_0\frac{\t_{00}^2+\t_{12}\t_{21}}{\t_{11}^2}+
\frac{\lp\t_{00}\rp_x\t_{21}-\t_{00}\lp\t_{21}\rp_x}{\t_{00}\t_{11}}.
\end{align}
\end{subequations}

\section{Vertex operators}
\label{sec:vertex}

We now introduce vertex operators that are eigenstates of
the nontrivial vacuum Lax operators \eqref{akns_vacuum}.
Consider $g$ in the usual solitonic form \eqref{g_soliton}, with
vertex operators defined by
\begin{subequations}\label{akns_vert_const}
\begin{align}
X_i &\equiv 
\sum_{n=-\infty}^{\infty} \biggl[
\dfrac{1}{2}\biggl(\k_i-\frac{q_0r_0}{\k_i}\biggr)\biggr]^{-n} X_i^{(n)} \\
& \quad \text{where} \quad X_i^{(n)} \equiv -\frac{q_0}{\k_i}\H{n}
+\frac{q_0\lp \k_i^2-q_0r_0\rp }{\k_i\lp \k_i^2+q_0r_0\rp }\delta_{n0}\hc
-\frac{q_0^2}{\k_i^2}\Ea{n}+\Eb{n}, \\
Y_i &\equiv 
\sum_{n=-\infty}^{\infty} \biggl[
\dfrac{1}{2}\biggl(\k_i-\frac{q_0r_0}{\k_i}\biggr)\biggr]^{-n} Y_i^{(n)} \\
&\quad \text{where} \quad Y_i^{(n)}\equiv -\frac{r_0}{\k_i}\H{n}
-\frac{r_0\lp \k_i^2-q_0r_0\rp }{\k_i\lp \k_i^2+q_0r_0\rp }\delta_{n0}\hc
+\Ea{n}-\frac{r_0^2}{\k_i^2}\Eb{n},
\end{align}
\end{subequations}
where $\k_i$ is a complex parameter.
In the limit $q_0,r_0\to 0$ these vertex operators recover
the well known vertices used for zero boundary condition \cite{ira}.
A direct calculation will give the following eigenvalue equations,
for $\O^{(n)}$ defined by \eqref{omega_n},
\begin{subequations}\label{akns_const_eigen}
\begin{align}
\lb\Omega^{(n)}, X_i \rb &= 
-\biggl[\dfrac{1}{2}\biggl( \k_i-\frac{q_0r_0}{\k_i}\biggr)\biggr]^{n-1}
\biggl[ \k_i+\frac{q_0r_0}{\k_i}\biggr] X_i, \\
\lb\Omega^{(n)}, Y_i \rb &= 
+\biggl[\dfrac{1}{2}\biggl( \k_i-\frac{q_0r_0}{\k_i}\biggr)\biggr]^{n-1}
\biggl[ \k_i+\frac{q_0r_0}{\k_i}\biggr] Y_i.
\end{align}
\end{subequations}
Therefore, the spacetime dependence of the solutions of system 
\eqref{akns-nvc} will be given by
\be\label{akns2_const_sp}
\begin{split}
\eta_i&= +\biggl( \k_i+\frac{q_0r_0}{\k_i}\biggr) x +
\frac{1}{2}\biggl( \k_i^2-\frac{(q_0r_0)^2}{\k_i^2}\biggr) t, \\
\xi_i&=  -\biggl( \k_i+\frac{q_0r_0}{\k_i}\biggr) x -
\frac{1}{2}\biggl( \k_i^2-\frac{(q_0r_0)^2}{\k_i^2}\biggr) t,
\end{split}
\ee
while for the system \eqref{akns3} we have
\be\label{akns3_const_sp}
\begin{split}
\eta_i&= +\biggl( \k_i+\frac{q_0r_0}{\k_i}\biggr) x +
\frac{1}{4}\biggl( \k_i^3-3\k_iq_0r_0-3\frac{(q_0r_0)^2}{\k_i}+
\frac{(q_0r_0)^3}{\k_i^3} \biggr) t, \\
\xi_i&= -\biggl( \k_i+\frac{q_0r_0}{\k_i}\biggr) x -
\frac{1}{4}\biggl( \k_i^3-3\k_iq_0r_0-3\frac{(q_0r_0)^2}{\k_i}+
\frac{(q_0r_0)^3}{\k_i^3} \biggr) t.
\end{split}
\ee
The matrix elements involving one vertex operator only are
\begin{subequations}\label{akns_const_vert1_matrix}
\begin{align}
\bra{\m_0}Y_i\ket{\m_0}&=-\frac{r_0\lp \k_i^2-q_0r_0\rp}
{\k_i\lp \k_i^2+q_0r_0\rp }, &
\bra{\mu_0}X_i\ket{\mu_0}&=\frac{q_0\lp \k_i^2-q_0r_0\rp }
{\k_i\lp \k_i^2+q_0r_0\rp }, \\
\bra{\mu_1}Y_i\ket{\mu_1}&= \frac{-2\k_ir_0}{\k_i^2+q_0r_0}, &
\bra{\mu_1}X_i\ket{\mu_1}&= \frac{-2q_0^2r_0}{\k_i\lp \k_i^2+q_0r_0\rp }, \\
\bra{\mu_1}Y_i\ket{\mu_2}&=1, &
\bra{\mu_1}X_i\ket{\mu_2}&=-\frac{q_0^2}{\k_i^2}, \\
\bra{\mu_2}Y_i\ket{\mu_1}&=-\frac{r_0^2}{\k_i^2}, &
\bra{\mu_2}X_i\ket{\mu_1}&=1.
\end{align}
\end{subequations}
Unlike the case of vanishing boundary condition, the solution
involving only one vertex is not trivial.
Choosing $g=e^{X_1}$ yields
\begin{subequations}\label{tau_1vertex}
\begin{align}
\tau_{00}&=1+\bra{\mu_0}X_1\ket{\mu_0} e^{\eta_1},\\
\tau_{11}&=1+\bra{\mu_1}X_1\ket{\mu_1}e^{\eta_1}, \\
\tau_{12}&=\bra{\mu_1}X_1\ket{\mu_2}e^{\eta_1},\\
\tau_{21}&=\bra{\mu_2}X_1\ket{\mu_1}e^{\eta_1}.
\end{align}
\end{subequations}
Analogous expressions are also valid for $Y_1$, replacing 
$\eta_1\to\xi_1$. Substituting the tau functions \eqref{tau_1vertex} in
\eqref{akns_fields_tau} we get the explicit 1-vertex solution of the positive
AKNS hierarchy.

For a solution involving two vertices like $g=e^{X_1}e^{Y_2}$, we have
more complicated matrix elements, but its general form is given by
\begin{subequations}
\begin{align}
\tau_{00} &= 1 + 
\bra{\m_0}X_1\ket{\m_0}e^{\eta_1}+
\bra{\m_0}Y_2\ket{\m_0}e^{\xi_2}+
\bra{\m_0}X_1Y_2\ket{\m_0}e^{\eta_1+\xi_2},\\
\tau_{11} &= 1 + 
\bra{\m_1}X_1\ket{\m_1}e^{\eta_1}+
\bra{\m_1}Y_2\ket{\m_1}e^{\xi_2}+
\bra{\m_1}X_1Y_2\ket{\m_1}e^{\eta_1+\xi_2},\\
\tau_{12} &=  
\bra{\m_1}X_1\ket{\m_2}e^{\eta_1}+
\bra{\m_1}Y_2\ket{\m_2}e^{\xi_2}+
\bra{\m_1}X_1Y_2\ket{\m_2}e^{\eta_1+\xi_2},\\
\tau_{21} &=  
\bra{\m_2}X_1\ket{\m_1}e^{\eta_1}+
\bra{\m_2}Y_2\ket{\m_1}e^{\xi_2}+
\bra{\m_2}X_1Y_2\ket{\m_1}e^{\eta_1+\xi_2}.
\end{align}
\end{subequations}
Besides the matrix elements involving one vertex operator
\eqref{akns_const_vert1_matrix}, the nilpotency property was explicitly
verified and the following two vertices matrix elements calculated,
\begin{subequations}\label{two_vertexes_matrix}
\begin{align}
\bra{\mu_0}X_iY_j\ket{\mu_0}&=\frac{
\lp \k_i^2-q_0r_0 \rp\lp \k_j^2-q_0r_0 \rp\lp \k_i\k_j+q_0r_0 \rp^2}{
\lp\k_i^2+q_0r_0\rp\lp\k_j^2+q_0r_0\rp\k_i\k_j\lp\k_i-\k_j\rp^2}, \\
\bra{\mu_1}X_iY_j\ket{\mu_1}&=\frac{
\lcurl q_0r_0\lp\k_i-\k_j\rp^2+\lp\k_i\k_j-q_0r_0\rp^2\rcurl
\lp\k_i\k_j+q_0r_0\rp^2}{
\lp\k_i^2+q_0r_0\rp\lp\k_j^2+q_0r_0\rp\k_i^2\lp\k_i-\k_j\rp^2},\\
\bra{\mu_1}X_iY_j\ket{\mu_2}&=-2q_0\frac{
\lp\k_i\k_j-q_0r_0\rp\lp\k_i\k_j+q_0r_0\rp^2}{
\lp\k_i^2+q_0r_0\rp\lp\k_j^2+q_0r_0\rp\lp\k_i-\k_j\rp\k_i^2},\\
\bra{\mu_2}X_iY_j\ket{\mu_1}&=-2r_0\frac{
\lp\k_i\k_j-q_0r_0\rp\lp\k_i\k_j+q_0r_0\rp^2}{
\lp\k_i^2+q_0r_0\rp\lp\k_j^2+q_0r_0\rp\lp\k_i-\k_j\rp\k_j^2}.
\end{align}
\end{subequations}
The matrix elements \eqref{two_vertexes_matrix} and 
\eqref{akns_const_vert1_matrix} recover known results when
$q_0,r_0\to 0$, corresponding to the vanishing boundary condition \cite{ira}.

\section{Conclusions}
\label{sec:conclusions}

We have shown how to employ the algebraic dressing method, using vertex
operators, to construct nonvaninshing (constant) boundary value
solutions for integrable hierarchies. This method was applied previously
for the mKdV hierarchy \cite{mkdv_even, gardner} and in this letter for 
the AKNS hierarchy, completing the possible classical models obtained
using the algebra $\hat{s\ell}_2$ with the construction 
of integrable hierarchies \cite{jfg}. 
We have algebraically explained how the dispersion relation and
multi-soliton interaction depend on the boundary conditions, here arising
as a consequence of the vertex operators \eref{akns_vert_const}.

Our solutions are quite general, solving at once various individual models
contained in the AKNS hierarchy like the focusing and defocusing NLS, 
mKdV, KdV and Gardner equations. We also have obtained an explicit 
2-soliton solution, involving two vertex operators. Of course higher
multi-soliton solutions can be constructed following the same approach, we
only face a technical difficulty in computing the matrix elements but the
procedure is well established.

\paragraph{Acknowledgements.} We thank CNPQ and Fapesp for financial 
suppport.  We would like to thank L. Feher for suggestions.



\end{document}